Generalized Ramsey Interferometry Explored with a Single Nuclear Spin Qudit

Clément Godfrin[1], Rafik Ballou[1], Edgar Bonet[1], Mario Ruben[2,3], Svetlana Klyatskaya[2], Wolfgang Wernsdorfer[1,2,4]*, Franck Balestro[1,5]*

[1] Univ. Grenoble Alpes, CNRS, Grenoble INP, Institut Néel, 38000 Grenoble, France.

[2] Institute of Nanotechnology, Karlsruhe Institute of Technology, 76344 Eggenstein-Leopoldshafen, Germany.

[3] Institute de Physique et Chimie de Matériaux (IPCMS), Université Strasbourg,23, rue du Loess, BP 43, 67034 Strasbourg cedex 2, France.

[4] Physikalisches Institut, Karlsruhe Institute of Technology (KIT), Wolfgang-Gaede-Str. 1, 76131 Karlsruhe, Germany

[5] Institut Universitaire de France, 103 boulevard Saint-Michel, 75005 Paris, France.

**Qu*d*its, with their state space of dimension *d* > 2, open fascinating experimental prospects[1]. The quantum properties of their states provide new potentialities for quantum information [2,3,4,5,6,7,8], quantum contextuality[9], expressions of geometric phases[10], facets of quantum entanglement[11] and many other foundational aspects of the quantum world, which are unapproachable with qu*b*its. We here experimentally investigate the quantum dynamics of a qudit (*d* = 4) that consists of a single 3/2 nuclear spin embedded in a molecular magnet transistor geometry, coherently driven by a microwave electric field. We propose and implement three protocols based on a generalization of the Ramsey interferometry to a multilevel system. First, the standard Ramsey interference is used to measure the accumulation of geometric phases. Then, two distinct transitions of the nuclear spin are addressed to measure the phase of an iSWAP quantum gate. Finally, through a succession of two Hadamard gates, the coherence time of a 3-state superposition is measured.**

A universal quantum computer requires the coherent control of large Hilbert spaces[12,13,14], which makes its achievement an ambitious technical challenge. The traditional approach is to build a large-scale quantum coherent architecture with two-states quantum devices (qu*b*its) into which information is encoded and treated using state populations and phases. An alternative path to overcome the scalability obstruction is to make use of *d*-states devices[15,16] (qu*d*its). Indeed, it was recently demonstrated that multi-level quantum systems can be of great relevance for the field of quantum information processing [2,3,4,5,6,7,8] provided a long coherent control of the phase of the system [17,18]. In general, interferometric circuits are used to get access to phase's information: an initial state is subjected to two paths before merging to a final state. Depending on the difference between the phases accumulated on the two different paths, the state will recombine in a constructive or destructive interference. The information on the phase difference between the two paths can be deduced from a population measurement. Furthermore, the phase coherence[19] can be assessed from the contrast of the interference fringes. The archetype interference experiment based on Young slits proved the wave/particle character of light[20] and single atoms[21]. Nowadays, Ramsey interferometry is widely used to characterize the coherence time of qu*b*its. We here propose a generalization of Ramsey interferometry to the case of a qu*d*it, which we have implemented in three different protocols. The first one (single-transition Ramsey interferometry) allows a phase measurement of quantum dynamics that preserves the state population. We use it to determine the geometric phase accumulated by quantum states driven along closed paths in their state spaces. The second protocol (double-transition Ramsey interferometry) is more general by demonstrating our ability to measure a phase of a quantum evolution even if state populations are modified. We apply it to the iSWAP quantum gate. The third protocol (Hadamard-Ramsey interferometry) is suited in essence to measure phases of multilevel-state superposition. We illustrate it here by measuring the coherence time of a three-state superposition. All these protocols can be generalized to the case of any *d*-state system and are universal. In order to highlight the feasibility and the potential of these protocols, they are illustrated by measurements performed on a single 3/2 nuclear spin located in a molecular magnet[7, 22, 23].

We exploit a bis(phthalocyanine)terbium (III) single-molecule magnet (TbPc$_2$), embedded in a single-electron transistor (figure 1(a)). The core of the molecule is a Tb$^{3+}$ ion with an electronic angular momentum $J = 6$ and a nuclear spin $I = 3/2$. Because of the ligand field of the Pc complexes, the electronic angular momentum behaves like a ±6 Ising spin. Its axis will be taken as quantization axis $z$ for the nuclear spin. The hyperfine interaction between the electronic and the nuclear spin consists of a dipolar term A ≈ 24.9 mK and a quadrupolar term $P$ ≈ 14.4 mK [24]. This specific hyperfine coupling gives rise to a four-level quantum system with three distinct resonance frequencies $v_1$ ≈ 2.45 GHz, $v_2$ ≈ 3.13 GHz and $v_3$ ≈ 3.81 GHz [7] (figure 1(b)). An exchange-coupling in between the Tb$^{3+}$ ion electronic spin and the spin carried by the Pc read-out quantum dot induces an electronic spin dependence of the conductance through the read-out quantum dot, allowing a direct read-out of this electronic spin *via* transport measurement[25]. Furthermore, this electronic spin has a finite probability to tunnel from one state to the other through Quantum Tunnelling of Magnetization (QTM) governed by the Landau-Zener process. Because of the hyperfine coupling, QTM can occur at four different magnetic fields, corresponding to the four nuclear spin states. Therefore, the magnetic field values at conductance jumps reads out the nuclear spin states[22]. Finally, an antenna in the vicinity of the transistor allows a coherent manipulation of the nuclear spin transitions using only electric field[7,23]. The device is cooled down to 40 mK by means of a dilution refrigerator and subjected to a 3D-magnetic field. The visibility of a given nuclear spin state, knowing the initial one, is then measured using the following protocol: (i) *Preparation*: magnetic field swept from +60 mT to -60 mT to read-out the initial state of the nuclear spin state. (ii) *Evolution*: application of electric microwave pulses at constant external magnetic field. (iii) *Reading-out*: magnetic field swept back from -60 mT to +60 mT to read-out the final nuclear spin state. The entire sequence is rejected when no QTM transition is detected. After repeating the procedure 500 times, we yield the visibility $V_{p,q}$ of the state $|q\rangle$ knowing that the initial state is $|p\rangle$ for a given pulse sequence $V_{p,q} = N_{p,q}/\sum_n N_{p,n}$, where $N_{p,q}$ counts the number of events for which the QTM reveals a state $|p\rangle$ before the pulse sequence and $|q\rangle$ after. The repetition of this sequence for different pulse lengths gives access to the dynamics of each state under the influence of the microwave pulse.

The microscopic mechanism by which the nuclear spin can be controlled with an electric field relies on the dependence of the hyperfine interactions on the electric field, which is amplified by the significant strengthening of the Stark interactions by the odd-parity contribution to the ligand field of the Pc complexes on the $Tb^{3+}$ ion causing a small mixing of electronic states of opposite parity[7, 23]. It follows from this hyperfine Stark effect that a time dependent electric field behaves with respect to the nuclear spin $I$ analogously to a magnetic field in rotation in the (*x,y*)-plane perpendicular to the electronic spin axis *z* with the same pulsations and phase shifts as the electric ones. Since the nuclear spin transverse operators $(I_x, I_y)$ connect each state $|M\rangle$ to the states $|M \pm 1\rangle$ the coherent control of the nuclear spin can be achieved through the three Rabi oscillations $|-3/2\rangle \leftrightarrow |-1/2\rangle$, $|-1/2\rangle \leftrightarrow |+1/2\rangle$ and $|+1/2\rangle \leftrightarrow |+3/2\rangle$, each taking place for a distinct pulsation (figure 1(b)). Whenever the manipulation of states is performed through a succession of monochromatic pulses the dynamics can be accounted for by making use of the Bloch sphere representation. The evolution after a time τ of an initial nuclear spin state $|\psi(0)\rangle$ subjected to a monochromatic pulse with the pulsation $\omega_{pq}$ of a $|p\rangle \leftrightarrow |q\rangle$ Rabi oscillation and a phase shift $\varphi$ can be described in the co-rotating frame by $|\psi(\omega_{pq}\tau)\rangle = R_\varphi^{|p\rangle,|q\rangle}(\omega_{pq}\tau)|\psi(0)\rangle$, with:

$$R_\varphi^{|p\rangle,|q\rangle}(\theta) = exp\left[i\,\theta\left(\cos\varphi\;\sigma_x^{|p\rangle,|q\rangle} + \sin\varphi\;\sigma_y^{|p\rangle,|q\rangle}\right)/2\right] \qquad (1)$$

where $(\sigma_x^{|p\rangle,|q\rangle}, \sigma_y^{|p\rangle,|q\rangle})$ operates on the space of states generated with $|p\rangle$ and $|q\rangle$ similarly than transverse Pauli operators in a spin-1/2 state space and cancels every state of the supplementary space. $R_\varphi^{|p\rangle,|q\rangle}(\theta)$ corresponds to a rotation of the angle $\theta$ about the axis at the angle $\varphi$ from the $x$ axis in the $(x,y)$-plane on the Bloch sphere $\Sigma_{pq}$ associated with the $|p\rangle \leftrightarrow |q\rangle$ Rabi oscillation. In the case where the manipulation of states is performed through a polychromatic pulse the dynamics can again be accounted for intuitively with rotations in a spin-$(n/2)$ state space for particular ratios of the amplitudes of the n chromatic components of the polychromatic pulse that can be experimentally calibrated. A transformation then is again denoted $R_\varphi^{|p\rangle,|q\rangle}(\theta)$ where the states $|p\rangle$ and $|q\rangle$ represents the extreme states of either a triplet (n=2) or the quadruplet (n=3).

**Single transition Ramsey interferometry: geometric phase.**

The phase of a state at any instant in a given protocol can be measured by generalizing the method of Ramsey interferometry as follows. Given a state $|p\rangle$, the two paths of an interferometer are built by creating a quantum superposition with the partner state $|q\rangle$ of a $|p\rangle \leftrightarrow |q\rangle$ Rabi oscillation. This is merely achieved by applying a $\pi/2$ pulse of pulsation $\omega_{pq}$ during the time $\tau = (\pi/2)/\omega_{pq}$. The states $|p\rangle$ and $|q\rangle$ then can be manipulated separately by microwave pulses of distinct pulsations what will feed them with independent phases $\varphi_p$ and $\varphi_q$. Now by applying the same $\pi/2$ pulse again the two arms of the interferometer are merged back what leads to a final state whose probability to be in the state $|p\rangle$ (resp. $|q\rangle$) is given by $\cos^2\left[\frac{\varphi_p - \varphi_q}{2}\right]$ (resp. $\sin^2\left[\frac{\varphi_p - \varphi_q}{2}\right]$) which reveals the difference in the phases separately accumulated by the two states from the time at which the first $\pi/2$ pulse was applied up to the time at which the second π/2 pulse was started. The computation of these probabilities is detailed in supplementary information. It is important to emphasize here that the population of the states $|p\rangle$ and $|q\rangle$ must be unaffected by the transformation that creates the phase difference in between the two paths. The motion of a quantum state on a closed path, which results in the accumulation of a geometric phase, ideally suited to this protocol. Geometric phases were first discovered by S. Pancharatnam[26] through polarization manipulation in classical optics. Later, M.V. Berry[27] found geometric phases by analysis of adiabatic cyclic quantum dynamics, then interpreted with a holonomy element in a line bundle[28], and finally discussed in its full generality in the context of non-abelian spectral bundles[10] and for non-cyclic and non-unitary evolutions[29]. Its global nature makes it robust to noise sources[30,31,32], what is a major interest for intrinsically fault-tolerant quantum information processing[33]. It was explored in various experiments including superconducting circuits[34,35] and molecular magnets[36]. In order to produce a geometric phase in the nuclear spin qudit we use a sequence of two consecutive π pulses of phase $\varphi$ and $\varphi + \Theta$ associated with a $|p\rangle \leftrightarrow |q\rangle$ transition, which corresponds to a closed path on a sphere composed of two meridian arcs shifted by an angle $\Theta$ and joining the north pole to the south pole. The geometric phase $\phi_G$ is given through the solid angle of the surface sustained by the closed path $\phi_G = |p - q|\Theta$ (see supplementary information). Note that the initial phase $\varphi$ is irrelevant and can be set arbitrarily to zero. In a first experiment we applied the following pulse sequence:

$$U_1 = \mathbf{R_0}^{|-3/2\rangle,|-1/2\rangle}(\boldsymbol{\pi/2}) \cdot R_\Theta^{|-1/2\rangle,|1/2\rangle}(\pi) \cdot R_0^{|-1/2\rangle,|1/2\rangle}(\pi) \cdot \mathbf{R_0}^{|-3/2\rangle,|-1/2\rangle}(\boldsymbol{\pi/2}) \quad (2)$$

The two end factors stand for the Ramsey interferometry. The geometric phase here is acquired by the state $|q\rangle = |-1/2\rangle$ when moving in the space of spaces associated with the transition $|q\rangle = |-1/2\rangle \leftrightarrow |r\rangle = |1/2\rangle$ isomorphic to a spin-1/2 state space. It is therefore equal to $\phi_G =$

$|q - r|\Theta = \Theta$. The $q$ arm of the interferometer is fed with the phase $\varphi_q = \phi_G$ whereas the $p$ arm remains unfed. It follows that $\varphi_p - \varphi_q = \Theta$ as observed experimentally in figure 3(a).

In a second experiment, the geometric phase was accumulated by the state $|q\rangle = |-1/2\rangle$ when moving in the space of spaces associated with $|-1/2\rangle \leftrightarrow |3/2\rangle$ transition isomorphic to a spin-1 state space:

$$U_2 = \mathbf{R_0}^{|-3/2\rangle,|-1/2\rangle}(\pi/2) \cdot \mathbf{R_\Theta}^{|-1/2\rangle,|3/2\rangle}(\pi) \cdot \mathbf{R_0}^{|-1/2\rangle,|3/2\rangle}(\pi) \cdot \mathbf{R_0}^{|-3/2\rangle,|-1/2\rangle}(\pi/2) \quad (3)$$

The $q$ arm of the interferometer is now fed with the phase $\varphi_q = \phi_G = |q - r|\Theta = 2\Theta$ whereas the $p$ arm remains unfed. It follows that $\varphi_p - \varphi_q = 2\Theta$ as again observed experimentally in figure 3(a).

In a last third experiment we aimed at measuring an interference pattern in between two geometric phases. In this purpose we applied the pulse sequence:

$$U_1 = \mathbf{R_0}^{|-1/2\rangle,|1/2\rangle}(\pi/2) \cdot V_{\Theta_1} \cdot V_{\Theta_3} \cdot \mathbf{R_0}^{|-1/2\rangle,|1/2\rangle}(\pi/2) \quad (4)$$

with $V_{\Theta_1} = \mathrm{R}_{\Theta_1}^{|-1/2\rangle,|-3/2\rangle}(\pi) \cdot \mathrm{R}_0^{|-1/2\rangle,|-3/2\rangle}(\pi)$ and $V_{\Theta_3} = \mathrm{R}_{\Theta_3}^{|1/2\rangle,|3/2\rangle}(\pi) \cdot \mathrm{R}_0^{|1/2\rangle,|3/2\rangle}(\pi)$. The two arms $|p\rangle = |-1/2\rangle$ and $|q\rangle = |1/2\rangle$ of the interferometer are now fed with a geometric phase, more precisely $\varphi_p = \Theta_1$ and $\varphi_q = \Theta_3$ from motions along closed paths in the state spaces associated with the transitions $|-1/2\rangle \leftrightarrow |-3/2\rangle$ and $|1/2\rangle \leftrightarrow |3/2\rangle$, respectively. The interference pattern in between the two geometric phases is achieved in the second Ramsey $\pi/2$ pulse as displayed in figure 3(c) where it is seen through the probability of being in the state $|q\rangle$.

**Double transition Ramsey interferometry: Quantum gate.**

Quantum operations on composite qubits can always be decomposed into a set of one-qubit gates and two-qubits gates and it proves of utmost interest to try out to implement the latter on a non-composite system in fact a single qudit, which thus can provide useful resources for quantum technologies[37]. We investigate in this purpose how a quartit enables an iSWAP gate. According to the following one-to-one map

$$|-3/2\rangle \leftrightarrow |0\rangle_A \otimes |0\rangle_B, \quad |-1/2\rangle \leftrightarrow |0\rangle_A \otimes |1\rangle_B, \quad |1/2\rangle \leftrightarrow |1\rangle_A \otimes |0\rangle_B, \quad |3/2\rangle \leftrightarrow |1\rangle_A \otimes |1\rangle_B$$

(5)

between the basis states of a quartit and those of 2 entangled qubits, one awaits from a quantum iSWAP gate that does not modify the states $|-3/2\rangle$ and $|3/2\rangle$ while it switches the states $|-1/2\rangle$ and $|1/2\rangle$ and feeds them with an additional $\pi/2$ phase without modifying the states $|-3/2\rangle$ and $|3/2\rangle$:

$$U_{iSWAP} = \begin{bmatrix} 1 & 0 & 0 & 0 \\ 0 & 0 & i & 0 \\ 0 & i & 0 & 0 \\ 0 & 0 & 0 & 1 \end{bmatrix} = R_0^{|-1/2\rangle,|1/2\rangle}(3\pi) \qquad (6)$$

As displayed in equation 6, an iSWAP quantum gate is implemented in a quartit through merely a resonant $3\pi$ pulse applied on the 2nd transition of the 4-level spectrum (figure 4(a)). However, the quantum gate affects both the phase and population of the states making it impossible to apply the previous Ramsey interferometry for the measurement of its phase. It is like attempting to measure the phase of an event that acts on both arms of the interferometer simultaneously. In order to circumvent this difficulty we propose to make use of a third arm in which we isolate one path, which is later used to interfere with the path modified by the implementation of the quantum gate. This is a more general Ramsey interferometry where the two π/2 pulses are not applied on the same transition. After an initialization of the system in the $|-1/2\rangle$ state, the two arms of the interferometer are created by applying a π/2 pulse on the $|-1/2\rangle \leftrightarrow |1/2\rangle$ second transition. Next, in order to preserve one path and to implement the quantum gate on the second one, a π pulse is applied on the $|1/2\rangle \leftrightarrow |3/2\rangle$ third transition then the second transition is driven during a duration τ. Finally, the two arms of the interferometer are merged back by applying a π/2 pulse on the $|-1/2\rangle \leftrightarrow |1/2\rangle$ second transition. The sequence is completed only for τ = π(1+2n) with n a integer number.

$$U_{iSWAP}^{Ramsey} = \mathbf{R_0}^{|1/2\rangle,|3/2\rangle}(\pi/2) \cdot \mathbf{R_0}^{|-1/2\rangle,|1/2\rangle}(\pi + 2n\pi)) \cdot \mathbf{R_0}^{|1/2\rangle,|3/2\rangle}(\pi) \cdot \mathbf{R_0}^{|-1/2\rangle,|1/2\rangle}(\pi/2)$$

(7)

Let us now focus on the output state as a function of n:

$$U_{iSWAP}^{Ramsey} \cdot \begin{pmatrix} 0 \\ 1 \\ 0 \\ 0 \end{pmatrix} = \begin{pmatrix} 0 \\ 0 \\ i\sin^2(n\pi/2) \\ -\cos^2(n\pi/2) \end{pmatrix} \quad (8)$$

Two cases are clearly distinguished:

- if *n* is even, what correspond to τ = π and an accumulated phase equal to –i, the output is the state $|3/2\rangle$

- if *n* is odd, what correspond to τ = 3π and an accumulated phase equal to i, the output is the state $|1/2\rangle$

The evolution of the different visibility as a function of the sequence time for a pulse equal to $\pi$ and $3\pi$ displayed in figure 4(b) and (c) respectively, experimentally prove the iSWAP quantum gate implementation and validate the selected strategy of the interferometry method.

**Hadamard-Ramsey interferometry: coherence time.**

Generally in quantum information protocol, Ramsey fringes are used to study incoherent process that affect a quantum superposition, thus measuring its coherence time. In the case of a qu*b*it a first $\pi/2$ pulse is applied to create a coherent superposition of states then the qu*b*it is let to evolve freely in its decohering environment before being projected back in the read-out basis via a second $\pi/2$ pulse. This is straightforwardly generalized to a qu*d*it. Instead of creating a 2-state superposition using a $\pi/2$ pulse, we apply a Hadamard gate[38] that creates a multi-level coherent superposition. The method will here be applied for a 3-state superposition ($|-3/2\rangle; |-1/2\rangle$ and $|1/2\rangle$). Since polychromatic pulses are used in order to drive several transitions simultaneously the Hamiltonian of the system in the ($|-3/2\rangle; |-1/2\rangle$ and $|1/2\rangle$) basis must be dealt with in the generalized rotating frame with respect to which it takes the form[7]:

$$H = \frac{\hbar}{2}\begin{pmatrix} 0 & \omega_1 & 0 \\ \omega_1 & 2\delta_1 & \omega_2 \\ 0 & \omega_2 & 2\delta_2 \end{pmatrix} \quad (9)$$

where $\delta_n$ are the pulsation detunings between the $n^{th}$ transition and the $\nu_n$ component of the pulse and where $\omega_n$ are the Rabi pulsation of the $n^{th}$ transition. The nuclear spin is initialized on the state $|-1/2\rangle$, then a first Hadamard gate is applied. It consists of a microwave pulse ensuring the same Rabi frequency for both resonances ($\omega = \omega_1 = \omega_2$) with a detuning $\delta_1 = \omega$ of duration $\tau_{Had} = \sqrt{3}\pi/3\omega$ (see supplementary information). Then the 3-state superposition is let under free evolution during a time τ and a second Hadamard gate is finally applied. The final state of the system after this sequence is the following:

$$U_{Had}\cdot W_\tau\cdot U_{Had}\cdot \begin{pmatrix}0\\1\\0\end{pmatrix} = \frac{1}{3}\begin{pmatrix}e^{i\omega\tau}-1\\e^{i\omega\tau}+2\\e^{i\omega\tau}-1\end{pmatrix} \qquad (10)$$

Where $U_{Had}$ is the Hadamard gate and $W_\tau$ is the free evolution of duration τ. In our case, the symmetry of the Hamiltonian ensures the same dynamic for the |-3/2⟩ and the |1/2⟩ states, thus reducing the dimensionality of the problem: all the necessary information recurs in the dynamic of the |-1/2⟩ state, which consists in oscillations at the Rabi pulsation $\omega$ in between the state |-1/2⟩ and the two states |1/2⟩ and |-3/2⟩ (figure 5(d)). The oscillation of the |-1/2⟩ state are displayed in figure 5(b) and (c) respectively for waiting time below 1 μs and above 25 μs. As for a qubit, incoherent interactions with the environment break the phase coherence causing a damping of the oscillation amplitudes (figure 5(e)). The characteristic time of this damping is in the order of 90 μs for this single nuclear spin qutrit. The method can be applied to any qudit system.

**Conclusion**

We investigated on a single nuclear spin qudit different interferometric protocols to measure geometric phase, quantum gate phase and finally multi-level quantum superposition coherence time. The periodicities of geometric phase accumulation, related to intrinsic properties of the Hilbert space under investigation, have been validated using Ramsey fringes. Then, through an analogy in between this system and 2-qubits, we demonstrate the implementation of an iSWAP gate. The phase of this gate can be directly measured using a multi-transition Ramsey fringes. Finally, using a protocol made of two Hadamard gate, we generalized the coherence time measurement of a two-level system to any n-level system. This protocol is illustrated by the measurement of a 3-state single nuclear spin superposition coherence time. These protocols and measurements open up a systematic way of quantum phase measurements on multi-level systems.

**Acknowledgments**

We gratefully acknowledge E. Eyraud, D. Lepoittevin, and C. Hoarau for their technical contributions and motivating discussions. We thank J.F. Motte, T. Crozes, B. Fernandez, S. Dufresnes and G. Julie for Nano-fabrication development, S. Thiele and C. Thirion for help with development of the experiment. Samples were fabricated in the NANOFAB facility of the Néel Institute. This work is partially supported by the French National Agency of Research through the ANR-13- BS10-0001 MolQuSpin, by the German Research Foundation (DFG) through the Transregio Project No. TR88 "3MET" and by the Alexander von Humboldt Foundation.

Correspondence and requests for materials should be addressed to:

wolfgang.wernsdorfer@neel.cnrs.fr and franck.balestro@grenoble.cnrs.fr

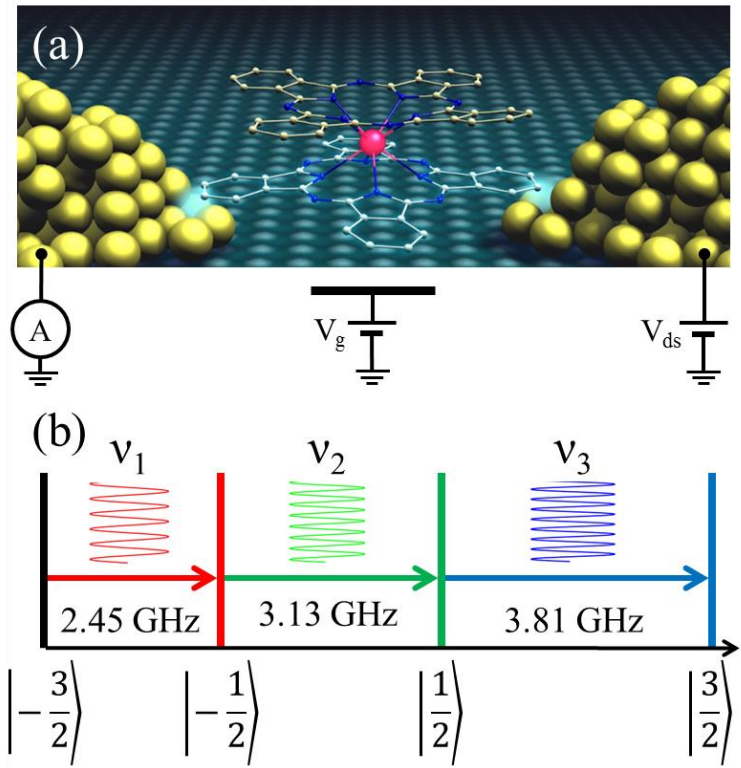

**Figure 1. Qudit Scheme |** (a) The TbPc₂ molecular magnet is embedded in a transistor. Transport measurement off the transistor under magnetic field enable the read-out of the 3/2 nuclear spin carried by the $Tb^{3+}$ ion. (b) Energy diagram of the four nuclear spin states. The quadrupole component in the hyperfine coupling enables an independent manipulation of each transition.

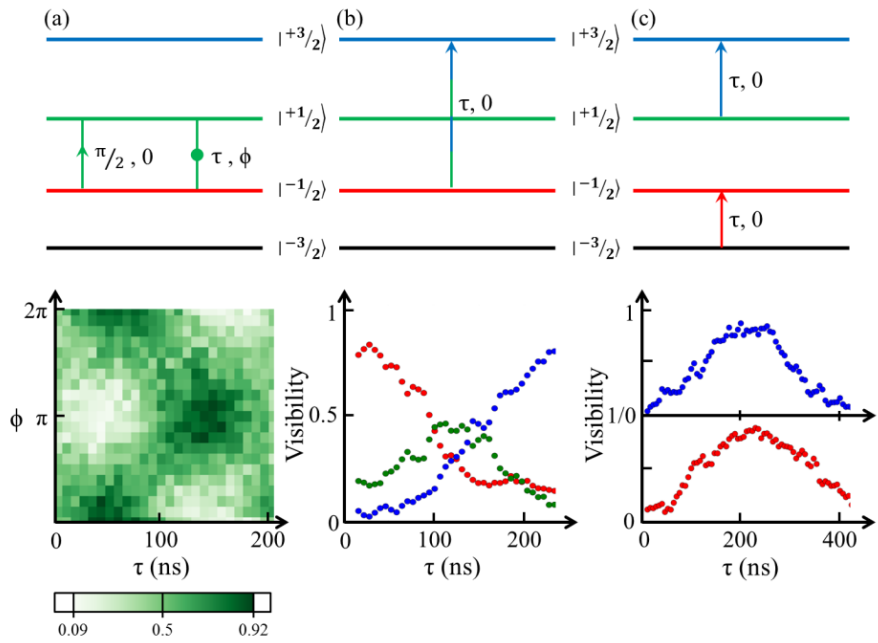

**Figure 2. Qubits coherent manipulation** | (a) A $\pi/2$ pulse with zero phase projects the state onto the equatorial plane of the Bloch-sphere. This first pulse is followed by a second one of duration $\tau$ with a phase $\phi$. The latter selects the rotation axis at the angle $\phi$ from the x-axis. The experimental measurement of the visibility as a function of $\tau$ and $\phi$ illustrates this dynamic. (b) A pulse comprising the frequency of the 2$^{nd}$ and the 3$^{rd}$ transition is sent during a time $\tau$. This pulse provides a population inversion in between the red and the blue state for a duration of 230 ns as observed in the evolution of each visibility as a function of $\tau$. It occurs through the intermediary of the green state. (c) A pulse comprising the frequency of the 1st and the 3$^{rd}$ transition is simultaneously sent during a time $\tau$. Visibility value of the red and the blue states starting from the black and the green one respectively is measured as the function of the pulse duration. Populations of these two states are reversed for pulse duration of 210 ns.

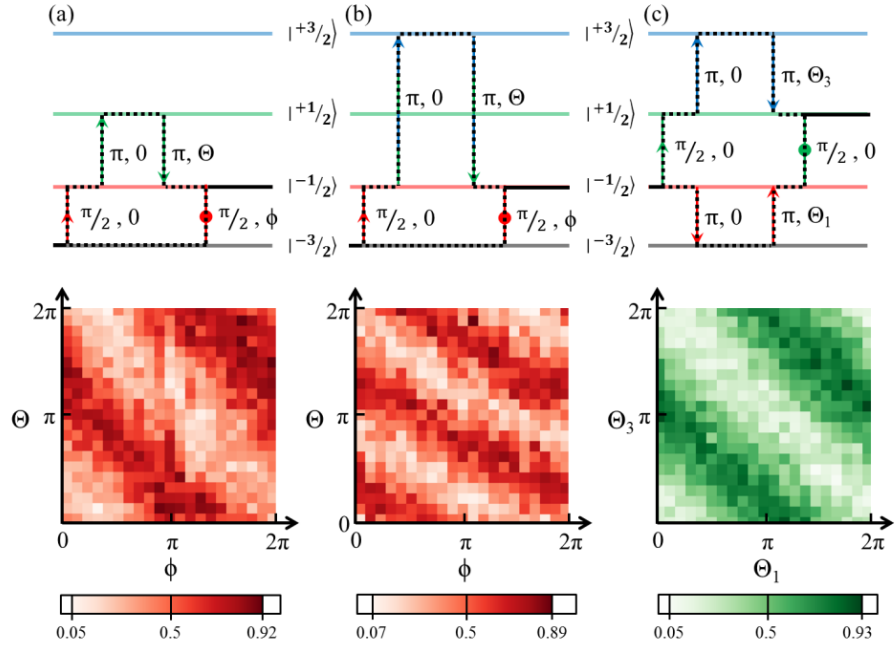

**Figure 3. Geometric phase** | (a) The system is initialized in the black state. A π/2 pulse on the first transition with zero phase creates a coherent superposition of the black and the red state. While no pulse is applied to the black state, a sequence of two π pulses is sent on the 2nd transition. The phase difference between these two pulses being Θ, a geometric phase equal to Θ is accumulated. Finally a second π/2 pulse is sent on the first transition with a phase ϕ. This creates $\cos^2 \frac{\Theta+\phi}{2}$ interference between the black and the red states. (b) The system is initialized in the black state. A π/2 pulse on the first transition with a zero phase creates a coherent superposition of the black and the red state. While no pulse is applied to the black state, a sequence of two π pulse is sent on the 2nd and the 3rd transition, as schematized in Fig. 2B. The phase difference between these two pulses being Θ, a geometric phase equal to 2Θ is accumulated. Finally a second π/2 pulse is sent on the first transition with a phase ϕ. It creates $\cos^2 \frac{2\Theta+\phi}{2}$ interference between the black and the red states. (c) The system is initialized in

the red state. A π/2 pulse on the second transition with a zero phase creates a coherent superposition of the red and the green state. Simultaneously a sequence of two π pulse is sent on the 1st and the 3rd transition, as schematized in Fig. 2C. The phase difference between these two pulse being $\Theta_1$ and $\Theta_3$, respectively. As a consequence geometric phases equal to $\Theta_1$ and $\Theta_3$ are accumulated on the red and on the green states respectively. Finally a second π/2 pulse on the first transition with a zero phase creates $\cos^2 \frac{\Theta_1+\Theta_3}{2}$ interference between the red and the green states. This interference pattern involves geometric phases only.

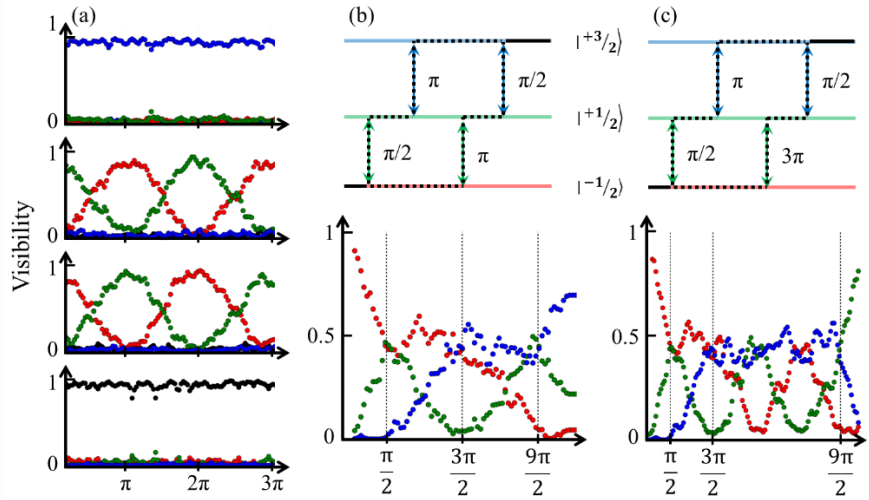

**Figure 4. iSWAP gate** | (a) A 3π pulse on the second transition with zero phase defines an iSWAP quantum gate. Visibility as a function of the pulse length of each state knowing that the initial state is $|-3/2\rangle, |-1/2\rangle, |1/2\rangle, |3/2\rangle$ from bottom to top respectively. The states $|-3/2\rangle$ and $|3/2\rangle$ are unchanged when the state $-|1/2\rangle$ and $|1/2\rangle$ are swapped. The 3π rotation ensures the accumulation of a dynamical phase equal to π. (b-c) To probe the phase "i = $e^{i\pi/2}$" of the state after the iSWAP gate we make use a three-arms Ramsey interferometry. The

peculiarity of the latter is that, in order to apply the quantum gate phase manipulation on only one arm of the interferometer, an additional π pulse is considered. Consequently, the π/2 pulses are sent on two different transitions. Visibility of each state as a function of the pulse length of the iSWAP quantum gate when the initial state is $|-3/2\rangle, |-1/2\rangle, |1/2\rangle, |3/2\rangle$ from bottom to top, respectively. With this sequence, the blue state visibility is maximized when the gate phase is equal to -1 and the green state visibility is maximized when the gate phase is equal to i.

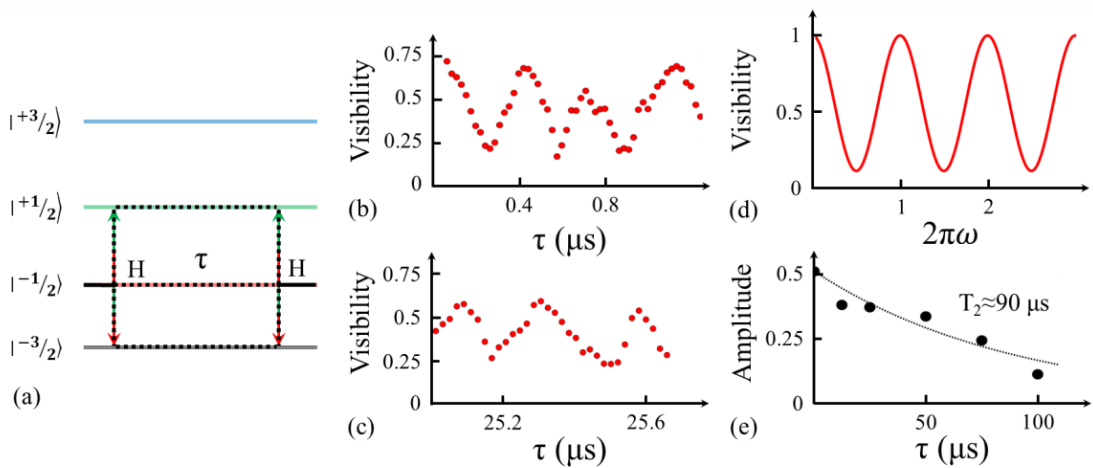

**Figure 5. 3-state coherence time** | (a) Pulse sequence to measure a multi-state superposition. A first Hadamard gate creates a coherent superposition, then the system evolves freely during a time τ. Finally a second Hadamard gate is sent to close the interferometric path. (b) Experimental oscillations of the red state as a function of the free evolution time. (c) The oscillations are still visible for τ of the order of 25μs. (d) Theoretical evolution of the red state visibility after applying two Hadamard gates separated in time by a free evolution time τ revealing a good agreement with the measurement. (e) Damping of the amplitude exhibits a coherence time for the 3 states superposition of the order of 90μs.

**Methods:**

The set-up is the same as that detailed in[23] excepting the microwave pulse generation. In this experiment the microwave pulse were generated with a Tektronix 24Giga sampling Arbitrary Wave Generator.